# Pure dephasing induced single-photon parametric down-conversion in an ultrastrong coupled plasmon-exciton system

Ruben Pompe[1], Matthias Hensen[2], Matthew Otten[3], Stephen K. Gray[4], and Walter Pfeiffer[1]

[1] *Fakultät für Physik, Universität Bielefeld, Universitätsstraße 25, 33615 Bielefeld, Germany*
[2] *Institut für Physikalische und Theoretische Chemie, Universität Würzburg, Am Hubland, 97074 Würzburg, Germany*
[3] *HRL Laboratories, LLC, Malibu, CA 90265, USA*
[4] *Center for Nanoscale Materials, Argonne National Laboratory, 9700 S. Cass Ave., Lemont, IL 60439, USA*

**Abstract**

In coupled quantum systems pure dephasing mechanisms acting on one constituent of the hybrid system break symmetry and enable optical transitions which are forbidden in the non-coupled system, i.e., the pure dephasing bath opens a cascaded dissipation pathway. Here we show that this mechanism enables single-photon induced parametric down-conversion in an ultrastrongly coupled plasmon-exciton system. Fast pure dephasing of the exciton is shown to support photon pair generation as the dominant energy relaxation pathway.

**PhySH:** Nanophotonics, Nonlinear optics, Cavity quantum electrodynamics, Open quantum systems & decoherence, Excitons, Hybrid quantum systems, Collective effects in quantum optics

Photon conversion via non-linear optical processes is a cornerstone in photonics. Besides the generation of coherent light at tunable wavelengths [1,2], it is key to generating nonclassical states of light [3,4] required for technologies like high precision interference measurements via squeezed light [5] or quantum encryption via entangled states [6]. Furthermore, nonlinear processes enable matter-mediated photon-photon interactions essential for optical information processing. Such nonlinearities provide ways to split and merge optical information channels or apply computational operations via optical switches. Nonlinear processes in optical crystals are a standard procedure for photon conversion [7]. Phase matching engineering and field concentration in photonic waveguides allows compensating their low nonlinear coefficients, thus increasing conversion efficiencies. These approaches work in the regime of many pump photons [8,9] with down-conversion efficiencies $\approx 4 \times 10^{-6}$ considered to be high [10], where the efficiency is defined as the probability to generate a down-converted photon pair per single incident photon. For applications in highly integrated optical circuits, such conversion efficiencies lead to high losses, which are detrimental to energy efficiency and make heat management necessary. Thus, optical nonlinearities occurring in the regime of single or few pump photons are of utmost importance [11]. Schemes using single photons and single atoms in a cavity at ultra-low temperatures reach the regime of single-photon nonlinearities [12]. Recently, room temperature single-photon nonlinearity was demonstrated using a Bose condensate of excitons in a microcavity [13]. Besides this many-emitter approach to achieve sufficient coupling, strong spatial confinement of electromagnetic modes in plasmonic nanoresonators also allows for strong coupling even with single quantum emitters [14,15] and offers an interesting route towards nanoscale nonlinear optics.

Here we show that ultrastrong coupling (USC) of a single two-level system (TLS) to a plasmonic resonator in combination with efficient pure dephasing of the TLS realizes a cyclic three-level system. A cyclic three-level system is the simplest quantum system to represent a $\chi^{(2)}$ material and can, in principle, achieve unit down-conversion efficiency [16–18]. The proposed scheme opens up ways to both reduce the size of a nonlinear optical device and achieve single-photon nonlinearities without major pumping losses at ambient conditions. Here, we focus on a down-conversion process, efficiently converting an incident photon into a photon pair.

The polariton, resulting from coupling of a boson mode to an excitonic TLS, provides the energy levels (Fig. 1a without bath interaction). The cavity quantum electrodynamics (cQED) Hamiltonian is [19]

$$H_{\text{sys}} = \underbrace{\Omega_m b^\dagger b}_{H_m} + \underbrace{\Omega_e \sigma_+\sigma_-}_{H_e} + \underbrace{g\,(b^\dagger + b)(\sigma_+ + \sigma_-)}_{H_{\text{int}}} + \underbrace{\frac{g^2}{\Omega_e}(b^\dagger + b)^2}_{H_{\text{dia}}}, \quad (1)$$

where $H_m$ and $H_e$ are the bare boson mode and TLS Hamiltonians, respectively. $b$ ($b^\dagger$) is the bosonic annihilation (creation) operator, $\sigma_-$ ($\sigma_+$) is the TLS lowering (raising) operator, and $\Omega_m$ and $\Omega_e$ are the boson mode and TLS level spacings, respectively. $H_{\text{int}}$ is the full dipole coupling Hamiltonian and $H_{\text{dia}}$ is the energy contribution from diamagnetic terms of the coupling with strength $g$. Inclusion of counter-rotating and diamagnetic terms is necessary in the USC regime, i.e. when $\eta = 2g/(\Omega_m + \Omega_e) \ll 1$ is not valid, since they significantly modify the energy spectrum [19]. The coupling results in the polariton eigenbasis $\{\text{GS}, \text{LP}, \text{UP}, \ldots\}$. The polariton ground state (GS), lower polariton state (LP) and upper polariton state (UP) provide the three-level system for parametric down-conversion (Fig. 1a, right panel), however, here a cascaded two-photon emission from UP is dipole forbidden.

Transitions in the system represented by Eq. (1) are treated in the Lindblad formalism [20] allowing for system-bath interactions and the interaction with a classical driving field. The temporal evolution of the system's density matrix $\rho_{\text{sys}}$ is then given by

$$\dot{\rho}_{\text{sys}} = \frac{-i}{\hbar}\left[H_{\text{sys}} + H_{\text{dr}}, \rho_{\text{sys}}\right] + \frac{1}{\hbar}\sum_i \Gamma_i\, L(o_i)\,\rho_{\text{sys}}, \quad (2)$$

with $H_{\text{sys}}$ as system Hamiltonian, $H_{\text{dr}}$ modeling external driving, and Lindblad terms $L(o_i)\rho = o_i \rho\, o_i^\dagger - (1/2)(\rho\, o_i^\dagger o_i + o_i^\dagger o_i\, \rho)$, where $o_i$ is the jump operator for the $i$-th system-bath interaction channel and $\Gamma_i$ is the corresponding coupling energy. In the case of uncoupled boson and exciton we consider three system-bath interaction channels, $i \in \{\text{m}, \text{e}, \varphi\}$. These are dissipation of the boson mode ($i = \text{m}$) with strength $\Gamma_m$ and jump operator $o_m = b$, exciton dissipation ($i = \text{e}$) with strength $\Gamma_e$ and jump operator $o_e = \sigma_-$, and exciton pure dephasing ($i = \varphi$) with jump operator $o_\varphi = \sigma_+\sigma_-$ and strength $\Gamma_\varphi$. For simplicity, pure dephasing of the boson mode, which for a plasmon could arise from electron-phonon interaction [21] or chemical damping [22], is neglected.

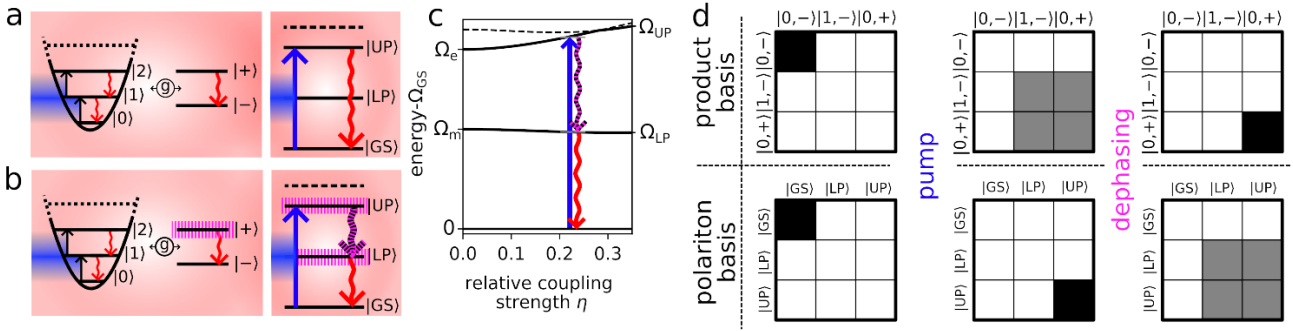

**FIG. 1. Pure dephasing induced down conversion in a polariton. a)** Externally driven (blue shading) boson mode with an eigenenergy indicated by black arrows coupled to an excitonic TLS are shown as separated subsystems in the left panel. Both subsystems are embedded in a bath (red background shading) which only supports dissipative transitions (red arrows). The equivalent eigenstate basis representation as a polariton is shown in the right panel. The upper polariton (UP) can be directly excited (blue arrow) and decays only directly back into the ground state (GS). Higher lying states are represented by horizontal black dashed lines. **b)** Same as a), but with additional pure dephasing acting on the TLS (purple shaded bar). In the polariton this results in an allowed UP to LP transition (purple/black dashed arrow), followed by an LP → GS transition, providing the photon down-conversion pathway. **c)** Polariton eigenenergies resulting from coupling of a detuned TLS and a boson mode ($\Omega_e/\Omega_m \approx 1.8$) as function of the relative coupling strength $\eta = 2g/(\Omega_e + \Omega_m)$. Arrows indicate the transition described for b) at the point of degenerate photon pair emission. The lower and higher energy photons are referred to as idler and signal, respectively. **d)** Schematic of the density matrix in the uncoupled product basis (upper row) and polariton eigenstate basis (lower row). First, the pump prepares the system in UP, an eigenstate of the polariton Hamiltonian. Next pure dephasing performs a measurement of the exciton state projecting the polariton into an eigenstate of $o_\varphi$, i.e., either $|0,+\rangle$ or $|1,-\rangle$ (here shown for $|0,+\rangle$), yielding LP population.

We consider the bosonic mode acting as an efficient nanoantenna with a dominating coupling to the driving field and negligible direct driving of the TLS. For weak coupling $g$ such external driving is mediated by the bosonic field operator $b^\dagger + b$. However, at sufficiently high coupling strength the boson-exciton hybridization can no longer be ignored and the driving term must be adapted. $H_{\text{dr}}$ is then [23]

$$H_{\text{dr}} = \kappa f(t)(X_b^\dagger + X_b), \qquad (3)$$

with $X_b = \sum_{j,k>j} \langle j|b^\dagger + b|k\rangle |j\rangle\langle k|$ in the polariton eigenbasis $j, k \in \{\text{GS, LP, UP}, ...\}$. $f(t)$ is a normalized real function reflecting the time-dependent driving and $\kappa$ is the effective coupling energy determined by the boson mode dipole moment and the maximum driving field. The same argument as for driving also holds for photon emission and the emission spectrum $S(\omega)$ is then obtained as the real part of the Fourier transform of the adapted field correlator, i.e., [24]

$$S(\omega) = \Gamma_m'/\hbar \int \int \text{Re}\{\langle X_b^\dagger(t) X_b(t+\tau)\rangle \exp(i\omega\tau)\} \, dt \, d\tau, \qquad (4)$$

where $\Gamma_m' = r_m \Gamma_m$ with $r_m$ being the ratio between the ratio between radiative loss strength $\Gamma_m'$ and total loss strength $\Gamma_m$ of the boson mode [25]. Again only photon emission via the boson mode is considered. Related quantities that require spectral selectivity, like photon pair yields, heralding efficiency, etc., are calculated via transition specific output observables (see Supplemental Material).

Treating bath interaction channels separately for each constituent also works only if the subsystems do not interact too strongly, i.e., $\eta \ll 1$. In the USC regime modifications are again necessary due to hybridization of the subsystems. In this case the full field and dipole coupling to the bath has to be considered for the dissipative jump operators for the subsystems, i.e., $o_m = b^\dagger + b$ and $o_e = \sigma_+ + \sigma_-$. Further, the jump operators of the coupled system $o_{jk} = |j\rangle\langle k|$ are expressed in the polariton eigenbasis and the transition strength is [26]

$$\Gamma_{jk} = \sum_i \Gamma_i |\langle j|o_i|k\rangle|^2 \qquad (5)$$

for each $k \to j$ transition, where $k \geq j$ holds for transition energies much larger than the thermal energy of the bath. The spectrum of the bath can have profound effects, e.g. Purcell enhancement [27] or suppression of photon emission in photonic band gaps [28], which can serve to further engineer effects. By employing the Lindblad formalism, we implicitly assume a bath with a white spectrum, i.e. approximating the system-bath coupling as flat within the considered spectral range.

Without pure dephasing of the TLS the UP relaxes back into the GS only via the emission of a single fluorescence photon (Fig. 1a). The cascaded emission of a photon pair is impossible since the transition matrix element for UP → LP vanishes, i.e., in Eq. (5) $\langle LP|o_m|UP\rangle = \langle LP|o_e|UP\rangle = 0$. In contrast, with pure dephasing $\langle LP|o_\varphi|UP\rangle \neq 0$ contributes to $\Gamma_{LP,UP}$, i.e. UP → LP transitions under emission of photons with energy $\Omega_{LP,UP} = \Omega_{UP} - \Omega_{LP}$ become allowed (Fig.1 b). When exclusively exciting UP these transitions feed the LP population and the emission of a second photon with energy $\Omega_{LP,GS} = \Omega_{LP} - \Omega_{GS}$ ($\langle GS|b^\dagger + b|LP\rangle \neq 0$, $\langle GS|\sigma_+ + \sigma_-|LP\rangle \neq 0$) is the consequence of the initial UP → LP transition, yielding correlated photon pairs. This process closes the three-step transition cycle of absorbing one photon and emitting two down-converted photons (Fig, 1b) serving as basis for our proposed scheme of single-photon parametric down-conversion. The boson to exciton detuning and as shown in Fig. 1c the coupling strength allow adjusting the transition energies. The strength of the UP → LP transition is determined by $\Gamma_{LP,UP} = \Gamma_\varphi |\langle LP|o_\varphi|UP\rangle|^2$ and depends on the pure dephasing strength $\Gamma_\varphi$ of the TLS and on the coupling strength $g$. For $\Gamma_{LP,UP} > \Gamma_{GS,UP}$ relaxation under emission of two cascaded photons becomes dominant and high down-conversion efficiencies are possible.

The finding that a pure dephasing mechanism opens new relaxation pathways for coupled systems can be rationalized based on the quantum trajectory formulation of system-bath interactions [20] (Fig. 1d): The system is initially in $|GS\rangle$, which - neglecting virtual photons - is approximated by the product state $|m, e\rangle = |0, -\rangle$ (no boson excitation and $-=$ TLS ground state). The pump excitation induces a GS → UP transition via $\langle UP|H_{dr}|GS\rangle \neq 0$. Resonant driving provides selectivity of the pump process, which is best understood in the polariton eigenbasis. In contrast, the pure dephasing mechanism acting on a subsystem is best conceived in the product state basis, where $|UP\rangle$ is a coherent superposition of $|0, +\rangle$ and $|1, -\rangle$. The pure dephasing jump operator $o_\varphi$ projects this superposition state either onto $|0, +\rangle$ or $|1, -\rangle$, i.e. an eigenstate of $o_\varphi$. However, both of these states are superposition states of $|UP\rangle$ and $|LP\rangle$ in the polariton eigenbasis. Hence, pure dephasing of a subsystem generates LP population and induces dissipation in the hybridized system causing emission of the first down-converted photon (Fig.1b).

The thus obtained cyclic three-level down-conversion scheme is universal and can be realized for any kind of coupled quantum system with at least one constituent subject to pure dephasing. USC favors photon pair emission and thus USC is the second key ingredient for our proposed scheme. USC has been achieved in a variety of systems, such as in circuits of superconducting qubits [29], intersubband polaritons [30], Landau polaritons [31], plasmonic modes and organic molecules [32], opto-mechanical systems [33] and plasmonic resonances coupled to a Fabry-Perot mode [34].

Focusing on the goal to realize down-conversion functionality in the VIS-IR range in nanoscale devices, we explore the above-proposed scheme for a semiconductor quantum dot (QD) exciton coupled to a tip-enhanced gap plasmon. Nano-star nanoparticles on a metal mirror substrate with dielectric spacing layer provide ultra-small mode volumes down to $V_m = 1.05 \times 10^{-7} \lambda_m^3$ ($\lambda_m = hc/\Omega_m$), and a dissipation strength of $\Gamma_m = 75$ meV [35] (Fig. 2a). To account for non-radiative losses of the plasmon we set $r_m = 0.5$ [25]. A semiconductor QD is used as the TLS because its properties are widely tunable, e.g. by altering geometry or material doping. Ultrafast dephasing for the TLS is desirable (cf. Figure 1d) and we assume the strongest pure dephasing strength for semiconductor QDs reported in literature of $\Gamma_\varphi = 94$ meV [36] as the upper limit. To reach the USC regime, we chose a high transition dipole moment $d = 140$ D [36], resulting in a dissipation strength of $\Gamma_e = 0.66$ meV. These parameters yield, based on the plasmonic field strength $E$ and the TLS dipole moment $d$, the maximal coupling strength $g_{max} = d|E| = d\sqrt{\Omega_m/(2\varepsilon_0 V_m)} > 0.5\, \Omega_m$. We limit the following examples to cases of $g < 0.3\, \Omega_m$ to account for, e.g., non-perfect positioning. For all simulations a transform limited Gaussian excitation pulse with an intensity full width at half maximum duration of 20 fs is used.

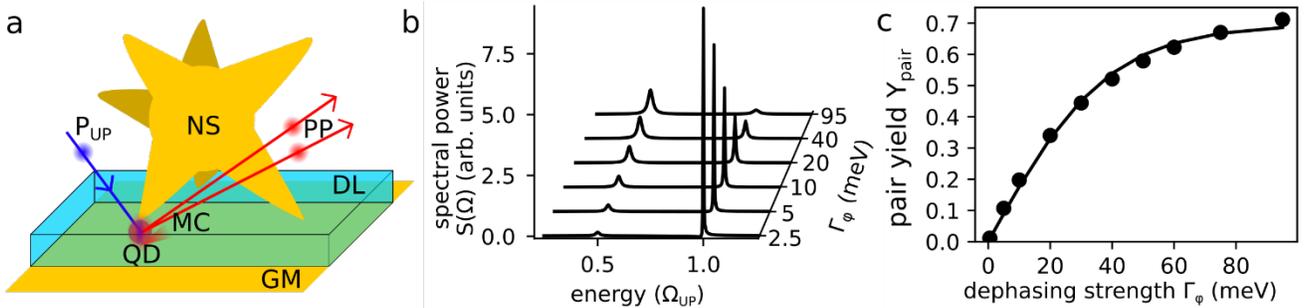

**FIG. 2. Dephasing induced degenerate down-conversion in an USC system. a)** Schematic of a possible realization of the proposed single-photon parametric down-conversion scheme. A gold nano-star (NS) placed on a gold mirror providing a mirror charge (MC) and separated by a few nm thick

dielectric layer (DL). Massive field enhancement (red shading) occurs between tips and mirror and thus in the dielectric layer [37], where a quantum dot (QD) is placed. As result, the QD is ultrastrongly coupled to the tip-enhanced gap-plasmon mode. A single pump photon (P$_{\text{UP}}$) is converted into a photon pair (PP). **b)** Emission spectra of the degenerate configuration for varying QD pure dephasing strength $\Gamma_\varphi$. Stronger pure dephasing increases down-conversion efficiency. Spectra are normalized to their respective total power. **c)** Photon pair yield per injected photon (dots) with sigmoid fit (curve). The plasmon resonance and exciton energy are $\Omega_m = 1.6$ eV ($\lambda_m = 775$ nm) and $\Omega_e = 2.9$ eV ($\lambda_e = 430$ nm), respectively. The relative coupling strength is $\eta = 0.233$ and the effective driving coupling is $\kappa = 260$ meV.

The first example is the efficient generation of photon pairs with equal signal and idler photon energy $\Omega_s = \Omega_i = 1.55$ eV (800 nm) from a $\Omega_p = 3.1$ eV ($\lambda_p = 400$ nm) pump photon. Depending on $\Gamma_\varphi$ pulsed excitation injects a mean total of $\langle N_{\text{in}}\rangle(\Gamma_\varphi = 0 \text{ meV}) = 0.97$ to $\langle N_{\text{in}}\rangle(\Gamma_\varphi = 94 \text{ meV}) = 0.77$ photons into the polariton system (for details see Supplemental Material). As seen in Fig. 2b for low and moderate pure dephasing of the TLS the direct pump photon fluorescence dominates. For stronger pure dephasing the photon pair emission starts to dominate the emission spectrum $S(\Omega)$. As shown in Fig. 2c for pure dephasing strengths $\Gamma_\varphi > 40$ meV the photon pair yield per injected photon $Y_{\text{pair}} > 0.5$. Losses are due to non-radiating deexcitations, UP → GS fluorescence, and, to a minor extent, spurious emission processes from excited states above UP (see Supplemental Material). The yield as function of the dephasing strength exhibits a saturation behavior (solid line in Fig. 2c). This is a direct consequence of the competing loss rates of the UP population. Because of the linear rise of $Y_{\text{pair}}$ for weaker pure dephasing even moderate pure dephasing yield already rather high down conversion efficiencies, before the saturation regime, limited by non-radiative losses, sets in.

To demonstrate a non-degenerate process, we show a configuration converting a pump photon at $\Omega_p = 1.94$ eV ($\lambda_p = 640$ nm) into a signal photon at $\Omega_s = 1.14$ eV ($\lambda_s = 1090$ nm) and an idler photon at $\Omega_i = 0.8$ eV ($\lambda_i = 1550$ nm) by adjusting detuning $\Omega_e - \Omega_m$ and coupling strength $g$ accordingly. The emission spectra (Fig. 3a) again reveal increasing down-conversion efficiency for increasing pure dephasing. For the given excitation the average number of injected photons decrease from $\langle N_{in}\rangle(\Gamma_\varphi = 0\text{ meV}) = 1.15$ to $\langle N_{in}\rangle(\Gamma_\varphi = 94\text{ meV}) = 0.75$. An injected photon results on average in 0.57 signal photons and 0.34 idler photons for $\Gamma_\varphi = 94$ meV. The loss mechanisms are the same as for the degenerate case. The signal-idler photon pair yield per injected photon can be again approximated by a sigmoid function reaching $Y_{pair}(\Gamma_\varphi = 94\text{ meV}) = 0.61$ (cf. Fig. 3b).

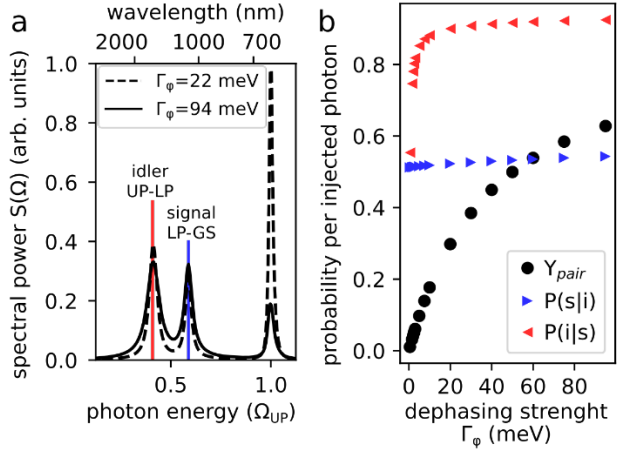

**Figure 3: Non-degenerate down conversion process for application in a nanoscale heralded single photon source. a)** Emission spectra (normalized to the respective total emission) as black solid line for $\Gamma_\varphi = 94$ meV, and dashed line for $\Gamma_\varphi = 22$ meV. **b)** Pair yield per injected photon (circle), confidence that a signal photon heralds an idler photon (right pointing triangle), and confidence that an idler photon heralds a signal photon (left pointing triangle) as function of the TLS pure dephasing strength $\Gamma_\varphi$. The nano-star gap-plasmon resonance is at $\Omega_m = 1.4$ eV ($\lambda_m = 900$ nm) and coupled with a relative strength of $\eta = 0.275$ to a QD with an exciton energy of $\Omega_e = 1.62$ eV ($\lambda_e = 765$ nm). The effective driving coupling is $\kappa = 120$ meV.

For such efficient down-conversion the scheme becomes interesting for heralded single photon sources. For $\Gamma_\varphi = 94$ meV the detection of a signal photon heralds an idler photon with 91% confidence, and the detection of an idler photon gives 53% confidence for a signal photon (Fig. 3b). The heralding capability of idler photons is rather independent of the dephasing, reflecting the fact that full LP population resulting from guaranteed idler emission inevitably leads to a LP → GS transition, radiating with assumed ~0.5 efficiency.

The device output efficiency including coupling losses is estimated by calculating the input coupling efficiency $\varepsilon$ as ratio between the diffraction limited spot size at the pump wavelength and experimental values for the extinction cross-section of nano-stars [38] with the respective resonance frequencies, resulting in $\varepsilon > 0.5$ for both, the degenerate and the non-degenerate case. We assume that the nano-star acts as antenna, directing all extinguished light into a single mode with sufficiently small mode volume. To account for deviations from these ideal cases $\varepsilon = 0.1$ is used to calculate output efficiencies of one pair per 15 illuminating photons for the degenerate and 24 for the non-degenerate case. This amounts in both cases to an increase in efficiency by four orders in magnitude in photon pair generation, compared to the highest experimental values [10].

To conclude, based on cQED calculations we propose an efficient single photon down-conversion scheme, realizable with existing technology. In this scheme, the phonon bath of a QD exciton provides pure dephasing for one constituent of a coupled quantum system and thus induces the necessary symmetry breaking that enables a cascaded photon pair emission. For ultrastrong coupling between a plasmonic nano-resonator and a semiconductor QD subjected to strong pure dephasing the pair production becomes the dominating relaxation pathway. This general scheme applies in any polaritonic system in which the UP-LP transition rate, enabled by the interplay of coupling and pure dephasing, is in the order of the UP-GS transition rate. Beyond ultrafast down-conversion this scheme offers means to decrease the impact of non-radiative losses and admits further non-linear functionalities like $\chi^{(3)}$ and $\chi^{(2)}$ photon up conversion. Notably, the scheme works also for resonators with rather high quality factors such as dielectric micro-resonators [39] and small coupling g, as long as the pure dephasing mechanism is strong enough.

This work was financially supported, in part, by the German Research Foundation (DFG) within the priority program SPP1839 grant PF317-11/1 (project # 410519108). Work performed at the Center for

Nanoscale Materials, a U.S. Department of Energy Office of Science User Facility, was supported by the U.S. DOE, Office of Basic Energy Sciences, under Contract No. DE-AC02-06CH11357.